\documentclass[twocolumn,prd,superscriptaddress,nofootinbib]{revtex4-2}

\usepackage{amsfonts,amsmath,amssymb,mathrsfs}
\usepackage{hyperref}
\usepackage{url}  
\usepackage{natbib}
\usepackage[mathletters]{ucs}
\usepackage[utf8x]{inputenc}
\usepackage{color}
\usepackage{graphicx}  
\usepackage{dcolumn}   
\usepackage{bm}        
\usepackage[english]{babel}
\usepackage[bottom]{footmisc}
\usepackage{booktabs,multirow,array}
\usepackage{dblfnote}
\usepackage{mathtools}
\usepackage{amsmath}
\usepackage{physics}
\DFNalwaysdouble 
\def\a{\alpha}

\def\r{\rho}
\def\s{\sigma}
\def\t{\tau}
\def\m{\mu}
\def\n{\nu}
\def\k{\kappa}
\def\th{\theta}
\def\g{\gamma}\def\G{\Gamma}
\def\L{t}\def\l{V}
\def\D{\Delta}
\def\la{\langle}
\def\ra{\rangle}
\def\o{\omega}\def\O{\Omega}
\def\d{\delta}
\def\p{\partial}

\def\oxthree{{\cal O}(x^3) }

\def\half{\textstyle{\frac{1}{2}}}
\def\bdoc{\begin{document}}
\def\edoc{\end{document}}
\def\bea{\begin{equation}}
\def\eea{\end{equation}}

\def\beq{\begin{eqnarray}}
\def\eeq{\end{eqnarray}}
\def\be{\begin{eqnarray}}
\def\ee{\end{eqnarray}}
\def\ben{\begin{enumerate}}
\def\een{\end{enumerate}}
\def\la{\langle}
\def\ra{\rangle}
\def\a{\alpha}
\def\g{\gamma}\def\G{\Gamma}
\def\d{\delta}\def\D{\Delta}
\def\e{\epsilon}
\def\z{\zeta}

\def\th{\theta}
\def\k{\kappa}
\def\l{t}
\def\m{\mu}
\def\n{\nu}
\def\o{\omega}
\def\p{\pi}
\def\r{\rho}
\def\s{\sigma}
\def\t{\tau}
\def\L{{\cal L}}
\def\S{\Sigma }
\def\gsim{\; \raisebox{-.8ex}{$\stackrel{\textstyle >}{\sim}$}\;}
\def\lsim{\; \raisebox{-.8ex}{$\stackrel{\textstyle <}{\sim}$}\;}
\def\gtrsim{\gsim}
\def\lessim{\lsim}
\def\loc{{\rm local}}
\def\vm{v_{\rm max}}
\def\bh{\bar{h}}
\def\del{\partial}
\def\nab{\nabla}
\def\half{{\textstyle{\frac{1}{2}}}}
\def\fourth{{\textstyle{\frac{1}{4}}}}

\def\bD{{\bf D}}
\def\bE{{\bf E}}
\def\bF{{\bf F}}
\def\bB{{\bf B}}
\def\bP{{\bf P}}
\def\bV{{\bf v}}
\def\bv{{\bf v}}
\def\bx{{\bf x}}
\def\by{{\bf y}}
\def\bz{{\bf z}}
\def\ba{{\bf a}}
\def\bd{{\bf d}}
\def\bs{{\bf s}}
\def\bn{{\bf n}}
\def\bp{{\bf p}}

\def\O{\Omega}

\def\br{{\bf r}}
\def\bnab{{\bf \nab}}

\def\tE{\tilde{E}}
\def\tL{\tilde{L}}
\def\Horava{Ho\v{r}ava }

\def\oxtwo{\mathscr{O}\left(x^2\right)}
\def\oxthree{\mathscr{O}\left(x^3\right)}
\def\oxfour{\mathscr{O}\left(x^4\right)}
\def\oxfive{\mathscr{O}\left(x^5\right)}
\def\LL{\text{Lanczos-Lovelock}}
\def\gr{\text{general relativity}}
\def\RN{Reissner-Nordstr\"{o}m}
\def\ph{\phantom}

\begin{document}
\title{
Overcharging Extremal Black Holes }
\author{Rajes Ghosh}
\email{rajes.ghosh@iitgn.ac.in }
\affiliation{Indian Institute of Technology, Gandhinagar, Gujarat 382355, India.}
\author{Akash K Mishra}
\email{akash.mishra@iitgn.ac.in }
\affiliation{Indian Institute of Technology, Gandhinagar, Gujarat 382355, India.}
\author{Sudipta Sarkar}
\email{sudiptas@iitgn.ac.in}
\affiliation{Indian Institute of Technology, Gandhinagar, Gujarat 382355, India.}

\begin{abstract}
The Weak Cosmic Censorship (WCC) conjecture can be used as a consistency criterion for any viable modification over general relativity (GR). We employ this idea to show that in contrast to the black holes in GR, it is indeed possible to overcharge modified extremal black hole solutions. Demanding the validity of WCC, we put some stringent constraints on various parameters of the theory. In particular, for charged Einstein-\ae ther black hole, our method is strong enough to reproduce the identical bound on the parameter obtained previously from entirely different considerations. 

 \end{abstract}

\maketitle

\section{Introduction and Motivation}\label{intro}
Our understanding of the contemporary gravitational physics is based on Einstein's theory of GR which describes the gravitational interaction in terms of the curvature of spacetime. In the presence of matter and/or energy, the spacetime metric is no longer flat, and it is determined by solving Einstein's field equations. Such a description of gravity in the language of geometry has been astonishingly successful in explaining various gravitational phenomena over a wide range of length scales. However, despite the success, GR has been confronted with several theoretical challenges over the years. One such severe limitation of GR is the existence of spacetime singularities occurring in the solutions of Einstein's equation. Hawking-Penrose singularity theorems~\cite{Penrose:1964wq,  Hawking:1966vg, hawking_ellis_1973, Wald:1984rg} ensure that the singularities arising due to geodesic incompleteness of causal curves are generic features of gravitational collapse. The basic notion of spacetime itself breaks down at the singularities, and physically relevant quantities become divergent. In such an apparently dreadful situation, Penrose's WCC conjecture~\cite{Penrose:1969pc} comes to rescue, which is essentially the statement that singularities that can causally influence the asymptotic regions of spacetime are censored out. More precisely, it negates the possibility of generating  \textit{naked singularity} as the end state of a gravitational collapse. The WCC is one of the most important conjectures in mathematical relativity, and it ensures the deterministic nature of the underlying theory of gravity. However, because of the complicated structure of the field equations, the conjecture still lacks a general proof \cite{Wald:1997wa,Joshi,Clarke_1994}. \\

In the absence of a general proof, the conjecture is often studied in the literature by analysing possible counterexamples. In these analyses, the general strategy is to start with an extremal or near extremal black hole solution as the initial configuration and look for any possible physical process that would lead to the formation of a naked singularity in the final state. If such a process can occur, it would be regarded as a potential violation of the WCC. This is first attempted in the seminal work of Wald~\cite{Wald}, and he proved that it is impossible to overcharge or overspin an extremal Kerr-Newman black hole via the process of test particle absorption. In followup works, the validity of WCC for several class of black hole spacetimes has been extensively studied~\cite{Hod:2008zza,Chirco:2010rq,BouhmadiLopez:2010vc,Saa:2011wq,Fairoos:2017lnm,Revelar:2017sem,Jana:2018knq,Shaymatov:2018fmp, Semiz:2005gs,Toth:2011ab,Nat_rio_2016,Jacobson:2009kt,Lehner:2010pn, Mishra:2019jsr}. Also, Hubeny showed that it is possible to overcharge a near extremal black hole to create a naked singularity in the final state via charged particle absorption as long as back-reaction effects can be neglected \cite{Hubeny:1998ga}. However, in a recent study~\cite{Sorce:2017dst}, the authors carried out a detailed analysis of the overcharging problem. They have proven by assuming the null energy condition on the matter that it is impossible to create naked singularities by overcharging extremal or near extremal black holes if the back-reaction effect is taken into account.  However, all these works are mainly focused on the black hole solutions of GR, and not much is known regarding the status of the WCC conjecture in alternative theories. For the particular case of charged black holes in Einstein-Gauss-Bonnet gravity, the WCC has been studied in Ref. \cite{Ghosh:2019dzq}.\\

Besides the issue of singularity discussed earlier, another critical limitation of GR is the incompatibility with quantum theory~\cite{Birrell:1982ix}. It is expected that at sufficiently small length scales, GR would be replaced by a quantum theory of gravity. These arguments strongly suggest that classical GR may only make sense as a low energy effective field theory which must be corrected appropriately at relevant length scales. These low energy quantum gravity corrections may manifest themselves in various forms in the effective Lagrangian. Results from string theory suggest that such corrections are often in the form of higher curvature invariants~\cite{Birrell:1982ix,PhysRevLett.55.2656,Zwiebach:1985uq}. Another special class of modifications over GR arises in the presence of local Lorentz violation. Although the local Lorentz invariance (LLI) is one of the fundamental guiding principles of GR, there are adequate reasons to suspect that Lorentz symmetry violation would occur at sufficiently high energy. In this context, the LLI may only be considered as an approximate symmetry of nature which is violated near the Planck scale. Signatures of such Lorentz violations can be incorporated in the effective gravitational theory by choosing a dynamical preferred frame, namely an \ae ther field. The resulting theory of gravity is known as Einstein-\ae ther theory, which is studied in ~\cite{PhysRevD.64.024028,Jacobson:2008aj}.\\

The current status of the validity of WCC in such alternative theories of gravity is not very clear. Therefore, in the absence of a general proof, it is crucial to understand the effect of such corrections on the cosmic censorship hypothesis. This motivates us to analyse the WCC by studying the overcharging problem via test particle absorption for a class of charged black hole solutions that differ from GR. The dynamical structure of these black hole spacetimes can be viewed as corrections over the \RN\ (RN) solution of GR arising due to the presence of any possible new physics, such as higher curvature terms, Lorentz violating vector fields etc. The strength of such corrections over and above GR are expected to be strongly constrained from both theoretical and observational considerations. One of the desirable features of any physical theory of nature should be a singularity-free description. As a result, the WCC can be used as a consistency criterion for any viable modification over GR.
Employing this idea, we show that in contrast to GR, it is possible to overcharge these modified extremal charged black holes unless some stringent bounds are put on various coupling constants present in the solution.\\

\section{The General Strategy}\label{str}

Let us consider a static, spherically symmetric and asymptotically flat black hole spacetime with mass $M$ and charge $Q$. The metric of such a spacetime is given below,

\bea \label{metric}
ds^2 = - f(r;\, M, Q)\, dt^2 + \frac{dr^2}{f(r;\, M, Q)} + r^2\, d\Omega^2\ ,
\eea

\noindent
with the electromagnetic vector potential, $A = - A_t(r)\, dt$. Although we only consider a four-dimensional spacetime, the following method can easily be generalized to higher dimensions as well. The term $d\Omega^2$ represents the metric on a unit $2$-sphere and the asymptotic flatness implies, $f(r;\, M, Q) \to 1$ as $r \to \infty$. Such a spacetime can occur as an exact solution of GR or any other higher curvature theory as well. We want to study the status of the WCC conjecture in this spacetime. In other words, if we start with an `extremal' black hole, whether it can be overcharged to a `naked singularity' by absorption of a test particle of charge $q$ and energy $E$.\\

Note that the locations of the horizons are given by the real positive roots of the `horizon equation': $f(r;\, M, Q)=0$, for a given mass $M$ and charge $Q$ of the central object. In general, there can be several positive roots of this equation which correspond to a spacetime with multiple horizons. However, for simplicity, we shall only consider those black hole solutions which have at most two horizons. Depending on the choice of the parameters $(M, Q)$, there are three possible scenarios--\\

(i) A black hole solution with two horizons. It occurs when the function $f(r;\, M, Q)$ has a negative value at its minimum $r=u$, i.e., $f(u;\, M, Q) < 0$.\\
\\(ii) An extremal black hole with one degenerate horizon. Such a situation arises when the horizon equation has a double root at $r=u$. It corresponds to `the extremal condition': $f(u;\, M, Q) = f'(u;\, M, Q) = 0$. Here, the prime denotes the derivative with respect to the radial coordinate.\\
\\(iii) A naked singularity with no horizon. It occurs when the function $f(r;\, M, Q)$ has a positive value at its minimum $r=u$, i.e., $f(u;\, M, Q) > 0$.\\

\noindent
Then, the set-up is as follows. We start with an extremal black hole with parameters $(M, Q)$ and throw a test particle of energy satisfying $0<E << M$ and charge $0< q << Q$ into it. We shall work under the `test particle approximation' and neglect the effect of back-reaction. Under this approximation, we can also neglect all the second-order terms like $q^2,\, E^2$, and $qE$ with respect to the first-order quantities. First, we want to derive `the overcharging condition', which sets an upper bound on the test particle's energy by demanding the final object with mass $(M+E)$ and charge $(Q+q)$ is a naked singularity. For this purpose, we Taylor expand the function $f(u+\epsilon;\, M+E, Q+q)$ about $(u;\, M, Q)$ and use the extremal condition. Here, the quantity $(u+\epsilon)$ gives the location where $f(r)$ is minimum after absorbing the test charge. Considering up to the linear order expansion in $\epsilon,\, E$, and $q$, we get the overcharging condition to be:

\bea \label{over}
E\,  < \, q\,. \left[\frac{\partial_Q f}{\abs{\partial_M f}}\right]_{(u;\, M, Q)}\ .
\eea
\noindent
Note that there are only three possible sign-structures of $\left[ \text{sgn}\left(\partial_M f(u)\right)\, ,\, \text{sgn}\left(\partial_Q f(u)\right) \right]$ compatible with the fact that the final state is a naked singularity. Among all of these, the cases of $[+,+]$ and $[+,-]$ give a lower bound on particle's energy (E) unlike Eq.(\ref{over}), and it is always possible to overcharge the initial extremal black hole satisfying the `entering condition' discussed below. Therefore, in any theory which has a spherically symmetric,  charged, asymptotically flat black hole solution in the form of Eq.(\ref{metric}), with  the sign-structure for $\left[ \text{sgn}\left(\partial_M f(u)\right)\, ,\, \text{sgn}\left(\partial_Q f(u)\right) \right]$ as $[+,+]$ and $[+,-]$, the WCC can always be violated using test particles. However, we consider only the third case with sign-structure $[-,+]$ as it provides an upper bound on energy E in contrast to the lower bound set by the `entering condition'. There is only a small window available for E obeying both bounds. Thus, in this case, there is a possibility of preserving WCC. However, to turn the extremal black hole into a naked singularity, it is necessary that the test charge crosses the horizon at $r=u$. This condition is dubbed as `the entering condition' and obtained by claiming the radial velocity of the test charge does not have any turning point before it reaches the horizon. It puts a lower bound on the particle's energy as follows:

\bea \label{enter}
E\, >\, q\, A_t(u)\, .
\eea

\noindent
For the specific case of a RN black hole, the extremal horizon is at $u = M = Q$, and the overcharging condition boils down to $E < q$. Moreover, the above entering condition as given in Eq.(\ref{enter}) reads $E > q$, which is exactly opposite of the overcharging condition. Thus, an extremal RN black hole can not be overcharged by test charge absorption\cite{Hubeny:1998ga}. Using a similar approach, one can also show that an extremal Einstein-Gauss-Bonnet charged black hole cannot be overcharged by test particle absorption \cite{Ghosh:2019dzq}. 
In general, the initial extremal black hole can be overcharged to turn into a naked singularity by test charge absorption iff there is a non-vanishing parameter space available for the energy ($E$) of the test charge. It is possible only when the following condition is satisfied,

\bea \label{naked}
A_t(u) < \left[\frac{\partial_Q f}{\abs{\partial_M f}}\right]_{(u;\, M, Q)} \ .
\eea

\noindent
When the above condition is satisfied, we can overcharge the extremal black hole and invalidate the WCC conjecture. Note that our analysis is only valid when the back-reaction effect is neglected, and we can use the test particle approximation. In this context, our goal is to consider various possible forms of the metric function $f(r, M, Q)$ and study if the extremal black hole can be overcharged. Interestingly, in the following section, we show that once we consider possible terms beyond GR, it is indeed possible to overcharge an extremal black hole by test charge absorption. This is in sharp contrast with the analogous result in GR. Moreover, by demanding the validity of WCC conjecture under test particle approximation, we can put some interesting constraints on various parameters of the solution as well.

\section{Testing WCC in modified black hole spacetime}\label{wcc}

In GR, the uniqueness theorems guarantee that the only static, asymptotically flat, electrovacuum black hole solution is RN. In the absence of any such theorem, the solution space of a modified theory is much larger. However, we only know very few exact black hole solutions of any modified gravity theory, particularly in $3+1$ spacetime dimensions. Therefore, we need to adopt a phenomenological approach to understand the overcharging problem for non-Einstein theories of gravity. For our study, we like to maintain the spherical symmetry and the asymptotic flatness of the solution. Then, we consider the following form of the metric function $f(r;\, M, Q)$ in Eq.(\ref{metric}),

\bea \label{f}
f(r;\, M, Q) = 1 - \frac{2M}{r} + \frac{Q^2}{r^2} + \sum_{k=3}^{N}\, \alpha_k \left( \frac{M}{r} \right)^k \ ,
\eea

\noindent
with the electromagnetic vector potential, $A = - Q/r\, dt$. The coupling coefficients $\alpha_k$'s are some dimensionless parameters. Note that the metric has a curvature singularity at $ r = 0$. This can be easily verified by calculating various curvature scalars. There are various ways to motivate such a spacetime metric. First of all, if we set all coefficients to zero except $\alpha_4$, the corresponding metric represents an exact charged solution of the Einstein-\ae ther theory \cite{Zhu:2019ura}. Similar solutions (without charge) are studied in the literature as Schwarzschild-like and Kerr-like cases \cite{JP}.\\

Moreover, such correction terms may also arise as Post-Newtonian (PN) corrections over GR~\cite{Will}.
Recently, such a parametrization is used to constraint the 2PN parameter from black hole shadow observations \cite{Psaltis:2020lvx}. Note that, in these references, the correction terms are assumed to be small so that the horizon structure remains the same except for some first-order corrections. However, we do not consider any such constraints in this work and treat all the coupling parameters non-perturbatively. We will show that the validity of the WCC puts strong constraints on the sign of these terms. \\

For convenience, we only consider spacetimes having at most two horizons. This implies, by using Descartes' rule of signs, all the coefficients $\alpha_k$ must be non-negative numbers. We choose the parameters $(M, Q)$ in such a way that the initial black hole is extremal, i.e., the function $f(r)$ has a doubly-degenerate root at $r=u$. This is tantamount to the condition $f(u;\, M, Q)= f'(u;\, M, Q)=0$, which gives `the extremal condition':

\bea \label{ex}
Mu-Q^2=\frac{1}{2}\, \sum_{k=3}^{N}\, \frac{k\, \alpha_k\, M^k}{u^{k-2}}\ .
\eea

\noindent
Note that for the RN case, all $\alpha_k$'s vanish, and we get $u=Q^2/M$. Substituting it back in the horizon equation, one gets the known extremal condition $M=Q$ for the RN black hole. \\

Now, consider the general case. Let us throw a test particle of charge $q$ and energy $E$ into the extremal black hole. Using Eq.(\ref{over}) and Eq.(\ref{f}), the overcharging condition reads as follows:

\bea \label{up}
E< \frac{q\, M}{Q}.
\eea

\noindent
Although this condition looks independent of the coefficients $\alpha_k$'s, it is not the case. This is because the extremal condition relates all the parameters of the theory, $M = M(Q,\, \alpha_k)$.\\

\noindent
The entering condition remains the same as given by Eq.(\ref{enter}). Thus, overcharging is possible if there is a non-vanishing parameter space available for $E$. It is possible only when Eq.(\ref{naked}) holds true, and for $A = - Q/r\, dt$ it reads,

\bea \label{uqm}
u > \frac{Q^2}{M}\ .
\eea

\noindent
This condition is indeed satisfied since by Eq.(\ref{ex}) the quantity $Mu-Q^2 > 0$, assuming $\alpha_k > 0$. It leads to the conclusion that when back-reaction effects are neglected, an extremal black hole can be overcharged by test particle absorption. Note that such a process is forbidden for extremal RN black holes.\\

In particular, the charged black hole of Einstein-\ae ther theory \cite{Zhu:2019ura, Ding} with only non-vanishing coefficient $\alpha_4=-c_{13}/(1-c_{13})$ can be overcharged if $\alpha_4 > 0$. 
If we demand that even with the test particle approximation, the WCC conjecture has to be satisfied as in \RN, we get the constraint: $0 \leq c_{13} < 1$. This bound on $c_{13}$ matches exactly with that given in \cite{Jacobson:2004ts, Berglund:2012bu} from completely different considerations, i.e., causality and stability of the perturbations. It is intriguing that the applicability of WCC also requires the same bound. Note that, when $\alpha_4 < 0$, the black hole spacetime always possesses at least one horizon for any range of $M$ and $Q$, which is ensured by Descartes' rule of signs. As a result, $\alpha_4 < 0$ turns out to be both necessary and sufficient condition for the validity of WCC.
\\

\noindent
If the metric under consideration is an exact solution of a higher curvature theory with dimensionful coupling $\beta_k$'s, then we can write the solution in a different form as

\bea \label{fh}
f(r;\, M, Q) = 1 - \frac{2M}{r} + \frac{Q^2}{r^2} + \sum_{k=3}^{N}\, \beta_k \frac{M^{k-2}}{r^k} \ .
\eea

\noindent
We may follow a similar procedure to obtain the condition for overcharging an extremal black hole of such theories. Again, we shall only consider the case when $\beta_k$'s are all non-negative to ensure that there is at most two horizons. It is straightforward to show that overcharging is possible in this case as well.\\

\noindent
Thus, when back-reaction is neglected, and test particle assumption is considered, extremal black holes with $f(r)$ of the form given by Eq.(\ref{f}) and Eq.(\ref{fh}) can be overcharged, unlike the RN case. Demanding the validity of WCC conjecture even at first-order puts stringent constraints on the $\alpha$ and $\beta$ coefficients of the theory.

\section{Discussions and conclusion}\label{con}
From the work of Hawking and Penrose, it has been long known that a sufficiently large amount of matter under gravitational collapse must lead to the formation of spacetime singularities. This set of results are known as singularity theorems. The existence of spacetime singularity in a generic solution of Einstein's equation leads to various pathological signatures. This includes the divergence of curvature invariants and the breakdown of physical laws. In this context, the WCC conjecture of Penrose is essential to ensure the deterministic nature of GR. In the absence of a general proof, the cosmic censorship hypothesis remains one of the major open problems in classical GR. \\
 
In this article, we studied the overcharging problem and consequently the validity of WCC in a particular class of charged black hole spacetime with a series of well-motivated corrections over GR. Such modifications may arise due to the presence of higher curvature terms or as PN corrections over GR. The form of spacetime metric we study also incorporates corrections due to local Lorentz violation. More precisely, the charged black hole solution of Einstein-\ae ther theory turns out to be a limiting case of our spacetime ansatz. We started by presenting a general analysis of the overcharging of extremal black holes via test particle absorption. Subsequently, we applied this methodology to study the validity of WCC in spacetimes given in Eq.(\ref{f}) and Eq.(\ref{fh}). In counter to the standard result that an extremal black hole can not be overcharged by test charge absorption, such possibilities are shown to be generic in some well-motivated modified theories unless the coupling constants obey certain strict bounds. In particular, our technique turns out to be powerful enough to reproduce the exact same bound on the coupling constant of a class of charged black hole solution in the Einstein-\ae ther theory. Also, given any exact solution, one may use our procedure to check the possibility of overcharging the extremal solution.\\

A possible generalization of our result would be to study the overspinning problem for a modified Kerr like solution. It may be possible to generate a rotating solution using Newman-Janis formalism~\cite{Newman:1965tw} from the metric in Eq.(\ref{metric}) and follow a similar procedure. Also, another interesting generalization would be to consider the possibility of non-minimal coupling between curvature and the electromagnetic field. Such a non-minimal coupling shall introduce new mass-dependent terms in the vector potential, and this may change the final conclusion.

\section*{Acknowledgement}
The research of RG is supported by the Prime Minister Research Fellowship (PMRF-192002-120), Government of India. The research of SS is supported by the Department of Science and Technology, Government of India under the SERB CRG Grant (CRG/2020/004562).



\begin{thebibliography}{100}




\section*{\bf{References}}    


\bibitem{Wald:1984rg}
R.  M.  Wald, General Relativity (Chicago  Univ.  Pr., Chicago, USA, 1984).


\bibitem{hawking_ellis_1973}
S. W. Hawking and G. F. R. Ellis,The Large Scale Structure of Space-Time, Cambridge Monographs on Mathematical Physics (Cambridge University Press, 1973).


\bibitem{Penrose:1964wq}
 R. Penrose, Gravitational collapse and spacetime singularities, Phys. Rev. Lett.14, 57 (1965).
 
 
\bibitem{Hawking:1966vg}
 S. W. Hawking, Singularities in the universe, Phys. Rev.Lett.17, 444 (1966).
 
 
 \bibitem{Penrose:1969pc}
  R.  Penrose,  Gravitational  collapse:  The  role  of  general relativity,  Riv.  Nuovo  Cim.1,  252  (1969),  [Gen.  Rel.Grav.34,1141(2002)].
  
  
\bibitem{Wald}  
R. .Wald, Gedanken experiments to destroy a black hole, Annals Phys.82, 548 (1974).


\bibitem{Wald:1997wa}
 R.  M.  Wald, Gravitational collapse  and  cosmic  censorship in black holes, Gravitational Radiation and the Universe: Essays in Honor of C.V. Vishveshwara(1997), arXiv:gr-qc/9710068 [gr-qc].


\bibitem{Joshi}
 P. S. Joshi, Cosmic censorship:  A current perspective,Modern Physics Letters A17, 1067 (2002).


\bibitem{Clarke_1994}
 C.  J.  S.  Clarke,  A  title  of  cosmic  censorship,  Classical and Quantum Gravity 11, 1375 (1994).
 

 
 
\bibitem{Hod:2008zza} 
 S.  Hod,  Weak  Cosmic  Censorship:  As  Strong  as  Ever, Phys.  Rev.  Lett.100,  121101  (2008),  arXiv:0805.3873[gr-qc].
 
 
\bibitem{Chirco:2010rq} 
 G. Chirco, S. Liberati, and T. P. Sotiriou, Gedanken experiments on nearly extremal black holes and the Third Law,  Phys.  Rev.D82,  104015  (2010),  arXiv:1006.3655[gr-qc].
 
 
\bibitem{BouhmadiLopez:2010vc} 
 M. Bouhmadi-Lopez, V. Cardoso, A. Nerozzi, and J. V.Rocha,  Black  holes  die  hard:   can  one  spin-up  a  blackhole past extremality?, Phys. Rev.D81, 084051 (2010),arXiv:1003.4295 [gr-qc].
 
 
\bibitem{Saa:2011wq}
A.  Saa  and  R.  Santarelli,  Destroying  a  near-extremal Kerr-Newman  black   hole,   Phys. Rev. D84,  027501(2011), arXiv:1105.3950 [gr-qc].




\bibitem{Fairoos:2017lnm}
C.  Fairoos,  A.  Ghosh,  and  S.  Sarkar,  Massless  charged particles: Cosmic  censorship,  and  the  third  law  of black  hole  mechanics,  Phys.  Rev.D96,  084013  (2017), arXiv:1709.05081 [gr-qc].


\bibitem{Revelar:2017sem}
K.   S.   Revelar   and   I.   Vega,   Overcharging   higher-dimensional black holes with point particles, Phys. Rev.D96, 064010 (2017), arXiv:1706.07190 [gr-qc].




 
 
\bibitem{Jana:2018knq} 
 S.  Jana,  R.  Shaikh,  and  S.  Sarkar,  Overcharging  blackholes and cosmic censorship in Born-Infeld gravity, Phys.Rev.D98, 124039 (2018), arXiv:1808.09656 [gr-qc].
 
 
 \bibitem{Shaymatov:2018fmp}
 S. Shaymatov, N. Dadhich, and B. Ahmedov, The higher dimensional Myers-Perry black hole with single rotation always obeys the Cosmic Censorship Conjecture,  (2018),arXiv:1809.10457 [gr-qc].
 
 
 
 
 
 
 \bibitem{Semiz:2005gs}
 I.  Semiz,  Dyonic  Kerr-Newman  black  holes,  complex scalar field and cosmic censorship, Gen. Rel. Grav.43,833 (2011), arXiv:gr-qc/0508011 [gr-qc].
 
 
 \bibitem{Toth:2011ab}
 G.  Z.  Toth,  Test  of  the  weak  cosmic  censorship  conjecture  with  a  charged  scalar  field  and  dyonic  Kerr-Newman black holes,  Gen. Rel. Grav.44, 2019  (2012),arXiv:1112.2382 [gr-qc].
 
 
 \bibitem{Nat_rio_2016}
 J. Natario, L. Queimada, and R. Vicente, Test fields can-not destroy extremal black holes, Classical and QuantumGravity33, 175002 (2016).

 




\bibitem{Jacobson:2009kt}
 T.  Jacobson  and  T.  P.  Sotiriou,  Over-spinning  a  blackhole  with  a  test  body,  Phys.  Rev.  Lett.103,  141101(2009),  [Erratum:   Phys.  Rev.  Lett.103,209903(2009)],arXiv:0907.4146 [gr-qc].
 
 
 \bibitem{Lehner:2010pn}
 L. Lehner and F. Pretorius, Black Strings, Low Viscosity Fluids, and Violation of Cosmic Censorship, Phys. Rev.Lett.105, 101102 (2010), arXiv:1006.5960 [hep-th].
 
 


\bibitem{Mishra:2019jsr}
 A. K. Mishra and S. Sarkar, Overcharging a multi-blackhole  system  and  cosmic  censorship,  Phys.  Rev.  D100,024030 (2019), arXiv:1905.00394 [gr-qc].
 
 
 
\bibitem{Hubeny:1998ga}
V.~E.~Hubeny,
Phys. Rev. D \textbf{59}, 064013 (1999)
doi:10.1103/PhysRevD.59.064013
[arXiv:gr-qc/9808043 [gr-qc]].\\



\bibitem{Sorce:2017dst}
J.~Sorce and R.~M.~Wald,
Phys. Rev. D \textbf{96}, no.10, 104014 (2017)
doi:10.1103/PhysRevD.96.104014
[arXiv:1707.05862 [gr-qc]].\\



\bibitem{Ghosh:2019dzq}
R.~Ghosh, C.~Fairoos and S.~Sarkar,
Phys. Rev. D \textbf{100}, no.12, 124019 (2019)
doi:10.1103/PhysRevD.100.124019
[arXiv:1906.08016 [gr-qc]].\\



\bibitem{Birrell:1982ix}
N.  D.  Birrell  and  P.  C.  W.  Davies, Quantum Fields in Curved Space, Cambridge Monographs on Mathematical Physics (Cambridge Univ. Press, Cambridge, UK, 1984).


\bibitem{PhysRevLett.55.2656}
D.  G.  Boulware  and  S.  Deser,  String-generated  gravity models, Phys. Rev. Lett.55, 2656 (1985).


\bibitem{Zwiebach:1985uq}
B. Zwiebach, Curvature Squared Terms and String The-ories, Phys. Lett. B156, 315 (1985).

 
 \bibitem{PhysRevD.64.024028}
 T. Jacobson and D. Mattingly, Gravity with a dynamical preferred frame, Phys. Rev. D64, 024028 (2001).
 
 
 
 
 \bibitem{Jacobson:2008aj}
 T.  Jacobson,  Einstein-aether  gravity:  A  Status  report,PoSQG-PH, 020 (2007), arXiv:0801.1547 [gr-qc].
 


\bibitem{Zhu:2019ura}
T.~Zhu, Q.~Wu, M.~Jamil and K.~Jusufi,
Phys. Rev. D \textbf{100}, no.4, 044055 (2019)
doi:10.1103/PhysRevD.100.044055
[arXiv:1906.05673 [gr-qc]].


\bibitem{JP}
T. Johannsen, D. Psaltis, Metric for rapidly spinning black holes suitable for strong-field tests of the no-hair theorem. Phys. Rev. D 83, 124015 (2011).



\bibitem{Will}
C. M. Will, Living Rev. Relativity 17, 4 (2014).



\bibitem{Psaltis:2020lvx}
D.~Psaltis \textit{et al.} [Event Horizon Telescope],
Phys. Rev. Lett. \textbf{125}, no.14, 141104 (2020)
doi:10.1103/PhysRevLett.125.141104
[arXiv:2010.01055 [gr-qc]].



\bibitem{Ding}
C. Ding, A. Wang, and X. Wang, Charged Einstein- aether black holes and Smarr formula, Phys. Rev. D 92, 084055 (2015).\\



\bibitem{Jacobson:2004ts}
T.~Jacobson and D.~Mattingly,
Phys. Rev. D \textbf{70}, 024003 (2004)
doi:10.1103/PhysRevD.70.024003
[arXiv:gr-qc/0402005 [gr-qc]].\\




\bibitem{Berglund:2012bu}
P.~Berglund, J.~Bhattacharyya and D.~Mattingly,
Phys. Rev. D \textbf{85}, 124019 (2012)
doi:10.1103/PhysRevD.85.124019
[arXiv:1202.4497 [hep-th]].\\


\bibitem{Newman:1965tw}
E.~T.~Newman and A.~I.~Janis,
J. Math. Phys. \textbf{6}, 915-917 (1965)
doi:10.1063/1.1704350








\end{thebibliography}
\end{document}